\documentstyle[aps]{revtex}
\begin{document}
\bibliographystyle{prsty}
\baselineskip=8.5mm
\parindent=7mm
\begin{center}
{\large {\bf \sc{ decay constants of the pseudoscalar  mesons in the framework of the coupled Schwinger-Dyson equation and
 Bethe-Salpeter equation}}} \\[2mm]
Zhi-Gang Wang$^{1,2}$ \footnote{E-mail,wangzgyiti@yahoo.com.cn.}, Wei-Min Yang$^{2,3}$ and Shao-Long Wan$^{2,3} $    \\
$^{1}$ Department of Physics, North China Electric Power University, Baoding 071003, P. R. China \footnote{Mailing address.}\\
$^{2}$ CCAST (World Laboratory), P.O.Box 8730, Beijing 100080,
P. R. China \\
$^{3}$ Department of Modern Physics, University of Science and Technology of China, Hefei 230026, P. R. China \\
\end{center}

\begin{abstract}
In this article, we investigate the structures of the pseudoscalar mesons ($\pi$, $K$, $D$, $D_s$, $B$ and $B_s$)
in the framework of
 the coupled rainbow Schwinger-Dyson equation and ladder
 Bethe-Salpeter equation with the confining effective potential (infrared modified flat bottom potential).
 The Schwinger-Dyson functions for the $u$, $d$ and $s$ quarks  are greatly renormalized  at small momentum region
and the curves are steep at about $q^2=1GeV^2$ which indicates an explicitly dynamical symmetry breaking.
The Euclidean time fourier transformed quark propagators have no mass poles in the time-like region which
 naturally implements  confinement.
As for the $c$ and $b$ quarks, the current masses are very large, the renormalization are more tender, however,
 mass poles in the time-like region are also absent. The Bethe-Salpeter wavefunctions for those mesons
 have the same type (Gaussian type) momentum dependence
 and center around small momentum which indicate that the bound states exist in the infrared region.
The decay constants for those  pseudoscalar mesons
are compatible with the values of experimental extractions and theoretical calculations, such as lattice simulations
 and QCD sum rules.
\end{abstract}


\section{Introduction}
The low energy nonperturbative properties of  quantum chromodynamics (QCD)
put forward a great challenge for physicists as the strong $SU(3)$ gauge coupling at small momentum region destroys
the perturbative expansion  approach, which is popular in theoretical physics.
The physicists propose many original  approaches
to deal with the long distance properties of QCD, such as Chiral perturbation theory \cite{Gasser85},
heavy quark effective theory \cite{Neubert94}, QCD sum rules  \cite{Shifman79}, lattice QCD \cite{Gupta98},
perturbative QCD \cite{Brodsky80}, coupled Schwinger-Dyson equation (SDE) and
Bethe-Salpeter equation (BSE) method \cite{Roberts94}, etc. All of those approaches have both outstanding advantages
and  obvious shortcomings in one or other ways.   The coupled rainbow SDE and ladder BSE have given a lot of successful
descriptions of the long distance properties of the low energy  QCD  and the QCD vacuum (for example, Refs.
\cite{DHL,MarisRoberts97,MarisTandy99,Ivanov99}
, for recent reviews one
can see Refs.\cite{Roberts00,Roberts03}). The SDE can naturally  embody the dynamical symmetry breaking and confinement
which are two crucial features of QCD, although they correspond to two very different energy scales \cite{Miransky93,Alkofer03}.
On the other hand, the BSE is a conventional approach in  dealing  with the two body
relativistic bound state problems \cite{BS51}. From the solutions of the BSE, we can obtain useful information
about the under-structure of the mesons and   obtain powerful tests for the quark theory.
However, the obviously drawback may be the model dependent kernels for the gluon two point Green's function
and the truncations for the coupled divergent SDE and BSE series in one or the other ways\cite{WYW03}.
Many analytical and numerical calculations indicate that
the coupled rainbow SDE and ladder BSE with phenomenological potential models can give model independent results and
 satisfactory values \cite{Roberts94,DHL,MarisRoberts97,MarisTandy99,Ivanov99,Roberts00,Roberts03}. The usually used
effective potential models are confining Dirac $\delta$ function potential,
Gaussian  distribution potential and flat bottom potential (FBP) \cite{Roberts00,Roberts03,Munczek83,Munczek91,Wangkl93}. The FBP is a sum of Yukawa
potentials, which not only  satisfies gauge invariance, chiral invariance and fully
relativistic covariance, but also suppresses the singular point which the
 Yukawa potential has. It works well in understanding the dynamical chiral symmetry braking, confinement and the QCD vacuum as well as
the meson  structures, such as electromagnetic form factors, radius, decay constants \cite{WangWan,Wan96,WYW03}.

The decay constants of the pseudoscalar mesons  play an important role in modern physics with the assumption
of current-meson duality. The precise knowledge of the those values (especially the $f_D$, $f_{D_s}$, $f_B$ and $f_{B_s}$) will provide
great  improvements  in our
understanding of various processes convolving the $D$, $D_s$, $B$ and $B_s$ mesons decays.
At present, it is a great challenge to extract
the values  of the $B$ and $B_s$ mesons decay constants $f_B$ and $f_{B_s}$ from the experimental data while the experimental
values for the $D$ and $D_s$ mesons decay constants $f_D$ and $f_{D_s}$  are plagued with large
uncertainties. So it is interesting to combine  those
successful  potentials within the framework of coupled SDE and BSE to calculate the decay constants of
the pseudoscalar mesons such as  $\pi$, $K$, $D$, $D_s$, $B$,  and $B_s$ while we prefer the detailed
studies of the quarkonium to another article. For previous studies about the electroweak decays of
the pseudoscalar mesons
with the SDE and BSE, one can consult Refs.  \cite{Roberts94,DHL,MarisRoberts97,MarisTandy99,Ivanov99,Roberts00,Roberts03}.
In this article, we use an infrared modified flat-bottom potential (IMFBP) which takes
 the advantages of
both the Gaussian distribution potential and the FBP to calculate the decay constants of those  pseudoscalar mesons.
The present article is a  continuation of our previous work \cite{WYW03}.

The article is arranged as follows:  we introduce the infrared modified flat bottom potential in section II;
in section III, IV and V, we solve the rainbow Schwinger-Dyson equation and  ladder Bethe-Salpeter
equation,  explore the analyticity of the quark propagators, investigate the dynamical symmetry breaking and confinement,
finally obtain the decay constants for those pseudoscalar  mesons; section VI is reserved for conclusion.

\section{Infrared modified Flat Bottom Potential }
The present techniques in QCD  calculation can not
give satisfactory large $r$ behavior for the gluon two point Green's function to implement
the linear potential
confinement  mechanism
\footnote{Here we correct a writing error in our article \cite{WYW03} i.e. change "small
$r$" to "large $r$"}, in practical manipulation, the phenomenological effective
 potential models always do the work.
As in our previous work \cite{WYW03}, we use
a gaussian distribution function to represent the infrared behavior of the gluon two point Green's function,
\begin{eqnarray}
4\pi G_{1}(k^2)=3\pi^2 \frac{\varpi^2}{\Delta^2}e^{-\frac{k^2}{\Delta}},
\end{eqnarray}
which determines the quark-antiquark interaction through a strength parameter $\varpi$ and a ranger parameter $\Delta$.
 This form is inspired by the $\delta$ function potential
(in other words the infrared dominated potential) used in Refs.\cite{Munczek83,Munczek91}, which it approaches in the limit
$\Delta\rightarrow 0$. For the intermediate momentum, we take the FBP as the best approximation and neglect
the large momentum contributions from the perturbative QCD calculations as the coupling constant at high energy
is very small.
The FBP is a sum of Yukawa potentials which is an analogy to the
exchange of a series of particles and ghosts with different
masses (Euclidean Form),
\begin{equation}
G_{2}(k^{2})=\sum_{j=0}^{n}
 \frac{a_{j}}{k^{2}+(N+j \rho)^{2}}  ,
\end{equation}
where $N$ stands for the minimum value of the masses, $\rho$ is their mass
difference, and $a_{j}$ is their relative coupling constant.
 Due to the particular condition we take for the FBP,
there is no divergence in solving the SDE.
In its three dimensional form, the FBP takes the following form:
\begin{equation}
V(r)=-\sum_{j=0}^{n}a_{j}\frac{{\rm e}^{-(N+j \rho)r}}{r}  .
\end{equation}
In order to suppress the singular point at $r=0$, we take the
following conditions:
\begin{eqnarray}
V(0)=constant, \nonumber \\
\frac{dV(0)}{dr}=\frac{d^{2}V(0)}{dr^{2}}=\cdot \cdot
\cdot=\frac{d^{n}V(0)} {dr^{n}}=0    .
\end{eqnarray}
So we can determine $a_{j}$ by solve the following
equations, inferred from the flat bottom condition Eq.(4),
\begin{eqnarray}
\sum_{j=0}^{n}a_{j}=0,\nonumber \\
\sum_{j=0}^{n}a_{j}(N+j \rho)=V(0),\nonumber \\
\sum_{j=0}^{n}a_{j}(N+j \rho)^{2}=0,\nonumber \\
\cdots \nonumber \\
\sum_{j=0}^{n}a_{j}(N+j \rho)^{n}=0 .
\end{eqnarray}
As in  previous literature \cite{Wangkl93,WangWan,Wan96,WYW03}, $n$ is set to be 9. The  phenomenological effective
 potential (infrared modified flat bottom potential) can be taken as $G(k^2)=G_1(k^2)+G_2(k^2)$.
\section{Schwinger-Dyson equation}
The Schwinger-Dyson equation can provide a natural
  framework for investigating the nonperturbative properties  of the
  quark and gluon Green's functions. By studying the evolution
  behavior and analytic structure of the dressed quark propagators,
  we can obtain valuable information about the dynamical symmetry
  breaking phenomenon and confinement.
 In the following, we write down the SDE for the quark propagator,
\begin{equation}
S^{-1}(p)=i\gamma \cdot p + m+\frac{16 \pi }{3}\int
\frac {d^{4}k}{(2 \pi)^{4}} \Gamma_{\mu}
S(k)\gamma_{\nu}G_{\mu \nu}(k-p),
\end{equation}
where
\begin{eqnarray}
S^{-1}(p)&=& i A(p^2)\gamma \cdot p+B(p^2)\equiv A(p^2)
[i\gamma \cdot p+m(p^2)], \\
G_{\mu \nu }(k)&=&(\delta_{\mu \nu}-\frac{k_{\mu}k_{\nu}}{k^2})G(k^2),
\end{eqnarray}
and $m$ stands for an explicit quark mass-breaking term. In this article, we take the rainbow approximation  $\Gamma_\mu=\gamma_\mu$.
With the explicit small mass term for the $u$ and $d$ quarks (comparing with the $u$ and $d$ quarks, the mass term for
 the $s$, $c$ and $b$ quarks  is large), we can preclude the zero
solution for the $B(p)$ and in fact there indeed exists a small bare current
quark mass. In this article, we take Landau gauge.
This dressing comprises the notation of constituent quark by
 providing a mass $m(p^2)=B(p^2)/A(p^2)$, for the $u$, $d$ and $s$ quarks, which is corresponding to
 the dynamical symmetry breaking. Because the form of
 the gluon propagator $G(p)$ in the infrared region can not be exactly inferred from the $SU(3)$ color gauge theory,
 one often uses model dependent forms as input parameters in the previous studies
  of the rainbow SDE
  \cite{Roberts94,Roberts00,Roberts03,Wangkl93,WangWan,Wan96,WYW03,Tandy97}, in this article we use the infrared modified FBP
to substitute for the gluon propagator.

In this article, we assume that a Wick rotation to Euclidean variables is
allowed, and perform a rotation analytically continuing $p$ and $k$
into the Euclidean region where they  can be denoted by $\bar{p}$ and $\bar{k}$,
 respectively.   Alternatively, one can derive the SDE from the
 Euclidean path-integral formulation of the theory, thus avoiding
 possible difficulties in performing the Wick
 rotation $\cite{Stainsby}$ . As far as only numerical results are concerned,
  the two procedures are equal. In fact, the analytical  structures of quark propagators have
 interesting information about confinement, we will make detailed discussion about
the $u$, $d$, $s$, $c$ and $b$ quarks propagators  respectively in  section V.

The Euclidean rainbow SDE can be projected into two coupled integral
equations for $A(\bar{p}^2)$ and $B(\bar{p}^2)$, the explicit expressions for those equations
can be found in Ref.\cite{WangWan,Wan96}. For simplicity, we
ignore the bar on $p$ and $k$ in the following notations.

\section{Bethe-Salpeter equation}
The BSE is a conventional approach in dealing with the two body
relativistic bound state problems \cite{BS51}. The quark theory of the mesons suppose that the mesons are
quark and antiquark bound states. The precise knowledge about the quark structures of the mesons
will result in better understanding of their properties. In the following, we write down the
ladder BSE for the pseudoscalar mesons,
\begin{eqnarray}
S^{-1}_{+}(q+\xi P)\chi(q,P)S^{-1}_{-}(q-(1-\xi)P)=\frac{16 \pi }{3} \int \frac{d^4 k}{(2\pi)^4}\gamma_\mu \chi(k,P)
\gamma_\nu G_{\mu \nu}(q-k),
\end{eqnarray}
where $S(q)$ is the quark propagator, $G_{\mu \nu}(k)$ is the gluon propagator, $P_\mu$ is the four-momentum
of the center of mass of the pseudoscalar mesons, $q_\mu$ is the relative four-momentum between the quark and antiquark
in the pseudoscalar mesons, $\gamma_{\mu}$ is the bare vertex of quark-gluon, $\xi$ is the center of mass parameter
which can be chosen to between $0$ and $1$, and
 $\chi(q,P)$ is the Bethe-Salpeter wavefunction (BSW) of the bound state.

We can perform the Wick rotation analytically and continue  $q$ and $k$ into the Euclidean region \footnote{To avoid possible
difficulties in performing the Wick rotation, one can derive the BSE from the Euclidean path-integral
formulation of the theory. As far as only numerical results are concerned,
  the two procedures are equal.},
the Euclidean pseudoscalar BSW $\chi(q,P)$ can be expanded in lorentz-invariant functions and  $SO(4)$ eigenfunctions,  Tchebychev polynomials
$T^{\frac{1}{2}}_{n}(\cos \theta)$. In the lowest order approximation, the BSW $\chi(q,P)$ takes the following form,
\begin{eqnarray}
\chi(q,P)=\gamma_5 \left[ iF_1^{0}(q,P)+\gamma \cdot P F_2^{0}(q,P)
+\gamma \cdot q q\cdot P F_3^{1}(q,P)+i[\gamma \cdot q,\gamma \cdot P  ] F_4^{0}(q,P) \right].
\end{eqnarray}
 It is important to translate the wavefunctions $F_{i}^{n}$ into the same mass dimension to facilitate the calculations
in solving the BSE,
\begin{eqnarray}
F_{1}^{0}\rightarrow F_{1}^{0}, \, F_{2}^{0}\rightarrow \Lambda^{1}F_{2}^{0}, \, F_{3}^{1}\rightarrow \Lambda^{3}F_{3}^{1},
\,F_{4}^{0}\rightarrow \Lambda^{2}F_{4}^{0}, \, q\rightarrow q/\Lambda, \, P\rightarrow P/\Lambda \, ,
\end{eqnarray}
where $\Lambda$ is some quantity of the dimension of mass.  Then
the ladder BSE can be projected into the following four coupled integral equations,
\begin{eqnarray}
H(1,1)F_1^0(q,P)+H(1,2)F_2^0(q,P)+H(1,3)F_3^1(q,P)+H(1,4)F_4^0(q,P)&=&\int_0^{\infty}k^3dk \int_0^{\pi}\sin^2 \theta K(1,1), \nonumber \\
H(2,1)F_1^0(q,P)+H(2,2)F_2^0(q,P)+H(2,3)F_3^1(q,P)+H(2,4)F_4^0(q,P)&=&\int_0^{\infty}k^3dk \int_0^{\pi}\sin^2 \theta (K(2,2)+K(2,3)), \nonumber \\
H(3,1)F_1^0(q,P)+H(3,2)F_2^0(q,P)+H(3,3)F_3^1(q,P)+H(3,4)F_4^0(q,P)&=&\int_0^{\infty}k^3dk \int_0^{\pi}\sin^2 \theta (K(3,2)+K(3,3)), \nonumber \\
H(4,1)F_1^0(q,P)+H(4,2)F_2^0(q,P)+H(4,3)F_3^1(q,P)+H(4,4)F_4^0(q,P)&=&\int_0^{\infty}k^3dk \int_0^{\pi}\sin^2 \theta K(4,4),
\end{eqnarray}
the expressions of the $H(i,j)$ and $K(i,j)$ are cumbersome and neglected here.

Here we will give some explanations for the expressions of $H(i,j)$ . The $H(i,j)$'s are functions of the quark's
Schwinger-Dyson functions (SDF) $A(q^2+\xi^2 P^2+\xi q \cdot P)$, $B(q^2+\xi^2 P^2+\xi q \cdot P)$,
 $A(q^2+(1-\xi)^2 P^2-(1-\xi) q \cdot P)$ and $B(q^2+(1-\xi)^2 P^2-(1-\xi) q \cdot P)$. The relative
four-momentum $q$ is a quantity in the Euclidean space-time  while the center of mass four-momentum $P$ is a quantity
 in the Minkowski space-time.
The present theoretical techniques  can not solve the SDE in the Minkowski space-time, we have to expand $A$ and $B$
in terms of Taylor series of  $q \cdot P$, for example,
\begin{eqnarray}
A(q^2+\xi^2 P^2+\xi q \cdot P)&=&A(q^2+\xi^2 P^2)+A(q^2+\xi^2 P^2)'\xi q \cdot P+\cdots. \nonumber
 \end{eqnarray}
The other problem is that we can not solve the SDE in the time-like region as the two
point gluon Green's function can not be exactly inferred from the $SU(3)$ color gauge theory
even in the low energy space-like region. In practical manipulations, we can extrapolate the values of $A$ and $B$
from the space-like region smoothly to the time-like region with suitable  polynomial functions.
To avoid possible violation with
confinement in sense of the appearance of pole masses $q^2=-m(q^2)$ in the time-like region, we must be care in
  choosing the polynomial functions \cite{Munczek91}.

Finally we write down the normalization condition for the BSW,
\begin{eqnarray}
\int \frac{d^4q}{(2\pi)^4} \{ \bar{\chi} \frac{\partial S^{-1}(q+\xi P)} {\partial P_{\mu}}\chi(q,P) S^{-1}(q-(1-\xi)P)
+\bar{\chi} S^{-1}(q+\xi P) \chi(q,P) \frac{\partial S^{-1}(q-(1-\xi) P)} {\partial P_{\mu}} \}=2 P_{\mu},
\end{eqnarray}
where $\bar{\chi}=\gamma_4 \chi^+ \gamma_4$. We can substitute the expressions of the BSWs and SDFs into the above
equation and obtain the analytical result, however, the expressions are cumbersome and  neglected here.

\section{Coupled rainbow SD equation and ladder BS equation and the decay constants}
In this  section, we explore the coupled equations of the rainbow SDE and ladder BSE for the pseudoscalar mesons.
In solving those equations numerically, the simultaneous iterations converge quickly to an unique value independent of
the choice of initial  wavefunctions. The final results for the SDFs and BSWs are plotted
as functions of the square momentum $q^2$.

 The quark-gluon vertex can be dressed through the solutions of the Ward-Takahashi identity or
 Salvnov-Taylor identity and taken to be the Ball-Chiu vertex and Curtis-Pennington vertex \cite{BallChiu,Curtis}. Although it is possible
to solve the SDE with the dressed vertex, our analytical results indicate that the expressions for the  BSEs with the dressed
vertex are cumbersome and not suitable for numerical iterations
\footnote{This observation is based on the authors's   work in USTC. }.

In order to demonstrate the confinement of quarks, we have to
study the analyticity of SDFs for the $u$, $d$, $s$, $c$ and $b$ quarks, and prove
 that there no mass poles on the real timelike  $q^2$ axial.
It is necessary to perform an analytical continuation of
the dressed quark propagators from the Euclidean space into the Minkowski
space $q_{4} \rightarrow iq_{0}$.
However, we have no knowledge about  the singularity structure of
quark propagators in the whole complex plane as our solutions are obtained in the Euclidean regions.
 We can take an
alternative safe procedure, stay completely in the Euclidean
space  and  take the Fourier transform
 with respect to the Euclidean time T
 for the scalar part ($S_{s}$) of the quark propagator \cite{Roberts94,Roberts00,Maris95},
 \begin{eqnarray}
 S^{*}_{s}(T) & & =\int_{-\infty}^{+ \infty} \frac{dq_{4}}{2 \pi}
 e^{iq_{4}T}S_{s}, \nonumber \\
 & & =  \int_{-\infty}^{+ \infty} \frac{dq_{4}}{2 \pi} e^{iq_{4}T}
 \frac{B(q^2)}{q^2A^2(q^2)+B^{2}(q^2)}|_{ \overrightarrow{q}=0},
 \end{eqnarray}
where the 3-vector part of $q$ is set to zero.
 If  S(q) has a mass pole at $q^2=-m^2(q^2)$ in the real timelike region, the Fourier transformed
  $S^{*}_{s}(T)$ would fall off as $e^{-mT}$ for large T or
  $\log{S^{*}_{s}}=-mT$.

In our numerical calculations, for small $T$, the values of $S^{*}_{s}$
are positive  and  decrease rapidly to zero and beyond with the increase  of $T$, which are compatible with the
result from  lattice simulations \cite{Bhagwat03}   ;  for large $T$, the values of $S^{*}_{s}$ are negative,
except occasionally a very small fraction positive values. We can express
$S^{*}_{s}$ as $|S^{*}_{s}|e^{i n\pi}$, $n$ is an odd integer,
$\log{S^{*}_{s}}=\log|{S^{*}_{s}}|+in \pi$.
If we neglect the imaginary part, we find that when the Euclidean time T is
large, there indeed exists a crudely approximated (almost flat) linear function with about zero slope for all the $u$, $d$
(in isospin symmetry limit, the u and d quarks are equal), $s$, $c$ and $b$ quarks with
respect to the variable T, which is shown in Fig.1. Here the word 'crudely' should be
understand in the linearly fitted sense, to be exact, there is no
linear function. However, such  crudely fitted linear functions are hard
to acquire  physical explanation and the negative values for $S^{*}_{s}$ indicate  an explicit
violation of the axiom of reflection positivity \cite{Jaffee},
 in other words, the quarks are not physical observable i.e. confinement.

From Fig.2, we can see that for the $u$,
$d$ (in isospin symmetry limit) and $s$ quarks, the SDFs are greatly renormalized at small momentum region and the curves are steep at about $q^2=1 GeV^2$ which
indicates  an explicit dynamical chiral symmetry breaking, while at large $q^{2}$,
they take asymptotic behavior.
As for the $c$ and $b$ quarks, shown in Fig.3, the current masses are very large, the renormalization is more tender,
however, mass poles  in the time-like region are also absent, which can be seen from Fig.1. At zero momentum,
$m_u(0)=688 MeV $, $m_d(0)= 688 MeV $, $m_s(0)= 882 MeV $, $m_c(0)= 1823 MeV $  and
$m_b(0)=4960 MeV $, which are compatible with the Euclidean constituent quark masses defined by $m^2(q^2)=q^2$
 in the literature.
 From the plotted BSWs (see Fig.4 as an example),
we can see that the BSWs for  pseudoscalar mesons have the same type (Gaussian type) momentum dependence while
   the quantitative values are different from each other. The gaussian type BS wavefunctions which
center around small momentum indicate that the bound states exist in the infrared region. Finally we obtain the values
for the decay constants of those  pseudoscalar mesons which are defined by
\begin{eqnarray}
i f_{\pi} P_\mu &=& \langle0|\bar{q}\gamma_\mu \gamma_5 q |\pi(P)\rangle, \nonumber \\
&=& \sqrt{N_c}\int Tr \left[\gamma_\mu \gamma_5\chi(k,P) \frac{d^4 k}{(2\pi)^4} \right],
\end{eqnarray}
here we use $\pi$ to represent the pseudoscalar mesons,
\begin{eqnarray}
f_\pi=127 MeV;  \, \, \, f_K=157 MeV; \, \, \, f_D=238 MeV; \, \, \,f_{D_s}=241 MeV; \,\,\, f_B=193 MeV; \,\,\,f_{B_s}=195 MeV,
\end{eqnarray}
which are compatible with the results from the experimental extractions, lattice simulations and QCD sum rules calculations,
$f_{\pi}=130MeV (Exp)$,
 $f_{K}=160MeV (Exp)$,
$f_{D}=300^{+180+80}_{-150-40}MeV (Exp)$,
$f_{D_s}=285\pm 19 \pm 40MeV (Exp)$,
$f_{D}\approx 160-230MeV (Latt, sumrule)$, $f_{D_s}\approx 210-260MeV (Latt, sumrule)$,
$f_{B}\approx 150-230MeV (Latt, sumrule)$ and
$f_{B_s}\approx 190-260MeV (Latt, sumrule)$ \cite{PDG02,Sumrule,Lattice}. The values from the
lattice simulations and QCD sum rules calculations vary  in a broad range, here we make an approximation.
In calculation, the input parameters are $N=1.0 \Lambda $, $V(0)=-11.0 \Lambda$,
 $\rho=5.0\Lambda$, $m_{u}=m_{d}=6 MeV$, $m_s=150MeV$, $m_c=1250MeV$, $m_b=4700 MeV$, $\Lambda=200 MeV$, $\varpi=1.6 GeV$ and  $\Delta=0.04 GeV^2$, the masses of the
pseudoscalar mesons are taken as input parameters \cite{WYW03}.

\section{conclusion }
In this article, we investigate the under-structures of the pseudoscalar mesons
($\pi$, $K$, $D$, $D_s$, $B$ and $B_s$)
in the framework of
the coupled rainbow SD equation and ladder BS equation with the confining effective potential
(infrared modified flat bottom potential).
Although the quark-gluon vertex can be dressed through the solutions of the Ward-Takahashi identity or
 Salvnov-Taylor identity and taken to be the Ball-Chiu vertex and Curtis-Pennington vertex,
a consistently numerical manipulation is unpractical, we take bare approximation for the vertex $\Gamma_\mu=\gamma_\mu$. After we solve the coupled rainbow SDE and ladder
BSE numerically, we obtain the SDFs and BSWs for the pseudoscalar mesons. The  SDFs for the $u$, $d$ and $s$
quarks are greatly renormalized  at small momentum region and the curves are steep at about $q^2=1 GeV^2$ which
indicates an explicitly dynamical chiral symmetry breaking. After we take the Euclidean time fourier transformation about
the quark propagators, we can find that there are no mass poles in the time-like region, and obtain satisfactory
results  about confinement. As for the $c$ and $b$ quarks, the current masses are very large,
the renormalization is more tender,
however,  mass poles in the time-like region are also absent. The BSWs for
the pseudoscalar mesons have the same type (Gaussian type) momentum dependence while
   the quantitative values are different from each other. The gaussian type BS wavefunctions which
center around small momentum indicate that the bound states exist  in the infrared region.
Our numerical results for the values of the
decay constants of the pseudoscalar are compatible with the corresponding ones obtained from
the experimental extractions   and other theoretical calculations, such as
lattice simulations  and QCD sum rules.
Once the satisfactory SDFs and BSWs for   the pseudoscalar mesons are known, we can use them to investigate a lot of important
quantities in the B meson decays, such as $B-\pi$, $B-K$, $B-D$, $B-\rho$ former factors, Isgur-Wise functions, strong coupling constants,
 etc.

\end{document}